\documentclass[preprint,aps]{revtex4}
\usepackage{graphicx}
\usepackage{amssymb}
\usepackage{epstopdf}
\usepackage{amsmath}
\usepackage{listings,textcomp}
\usepackage[usenames, dvipsnames]{color}
\usepackage[ pdftex, plainpages = false, pdfpagelabels, 
                 pdfpagelayout = useoutlines,
                 bookmarks,
                 bookmarksopen = true,
                 bookmarksnumbered = true,
                 breaklinks = true,
                 linktocpage=all,
                 pagebackref=false,
                 colorlinks = true,
                 linkcolor = BrickRed,
                 urlcolor  = blue,
                 citecolor = BrickRed,
                 anchorcolor = green,
                 hyperindex = true,
                 hyperfigures
                 ]{hyperref}

\newcounter{figref}

\begin{document}

\title{\href{http://www.necsi.edu/research/social/southafrica/}{South African Riots: \\ Repercussion of the Global Food Crisis and US drought}}
\date{January 8, 2013}
\author{Yavni Bar-\!Yam, Marco Lagi and \href{http://necsi.edu/faculty/bar-yam.html}{Yaneer Bar-\!Yam}}
\affiliation{\href{http://www.necsi.edu}{New England Complex Systems Institute} \\ 
238 Main St. Suite 319 Cambridge, MA 02142, USA \vspace{2ex}}

\begin{abstract}
High and volatile global food prices have led to food riots and played a critical role in triggering the Arab Spring revolutions in recent years. The severe drought in the US in the summer of 2012 led to a new increase in food prices. Through the fall, they remained at a threshold above which the riots and revolutions had predominantly occurred. Global prices at this level create conditions where an exacerbating local circumstance can trigger unrest. Global corn (maize) prices reached new highs, and countries that depend mostly on maize are more likely to experience high local food prices and associated pressures toward social unrest. Here we analyze the conditions in South Africa, which is a heavily maize-dependent country, and we find that consumer food indices have increased dramatically. Coinciding with the food price increases this summer, massive labor strikes in mining and agriculture have led to the greatest single incident of social violence since the fall of apartheid in 1994. Worker demands for dramatic pay increases reflect that their wages have not kept up with drastic increases in the prices of necessities, especially food. Without attention to the global food price situation, more incidents of food-based social instability are likely to arise. Other countries that have manifested food-related protests and riots in 2012 include Haiti (prior to the impact of Hurricane Sandy) and Argentina. Moreover, these cases of unrest are just the most visible symptom of widespread suffering of poor populations worldwide due to elevated food prices. Our analysis has shown that policy decisions that would directly impact food prices are decreasing the conversion of maize to ethanol in the US, and reimposing regulations on commodity futures markets to prevent excessive speculation, which we have shown causes bubbles and crashes in these markets. Absent such policy actions, governments and companies should track and mitigate the impact of high and volatile food prices on citizens and employees. 
\end{abstract}

\maketitle

On August 2, 2012, a violent riot at a platinum mine in South Africa resulted in three deaths \cite{earliest}. Subsequent events throughout the platinum mining region included a particularly violent incident on August 16 at the Marikana mine, which resulted in  34 strikers killed and about 80 more injured---the most violent such incident since the end of apartheid in 1994 \cite{most_violent,massacre,ABC}. The protests were taken up by gold miners \cite{gold} and agricultural workers, resulting in the destruction of over 120 acres of crops \cite{vineyards}. Observers considered these incidents to be so severe that they may undermine the country's peaceful reputation \cite{massacre,ABC}. The only event of comparable magnitude since the end of apartheid was a wave of xenophobic violence in May of 2008. 

The conditions that give rise to social violence are often poorly understood. Here, however, we provide evidence for a link between the violence in South Africa and the rapidly rising food prices that have affected many parts of the world, including triggering widespread food riots in 2007-08 and the uprisings in North Africa and the Middle East often called the Arab Spring in 2010-11 \cite{food_crises}. At the beginning of August 2012, when the labor riots started, corn (maize) prices rose to record highs driven in part by a drought in the US Midwest, but also by other underlying causes that have increased food prices rapidly in recent years. The earlier xenophobic violence \cite{xenophobia} coincided with the 2007-08 global food price peak that is linked to food riots in 30 countries.

In earlier papers \cite{food_crises,food_prices,Feb_update,July_update,food_for_fuel,food_faq} we have shown
that high and spiking global food prices have triggered riots and revolutions, and warned that predicted future spikes in global food prices would lead to additional social unrest \cite{food_crises}. We constructed a quantitative model that identified the causes of the rises in global food prices to be biofuels, particularly the conversion of corn to ethanol in the U.S., and increased financial speculation in the agricultural commodities futures market \cite{food_prices}. In addition to the global analyses, we have analyzed the origins of violence in Yemen \cite{yemen} and the increase in cost of Mexico's maize imports \cite{mexico}.

\begin{figure}[tb]
\centering
\refstepcounter{figref}\label{fig1}
	\includegraphics[width=150mm]{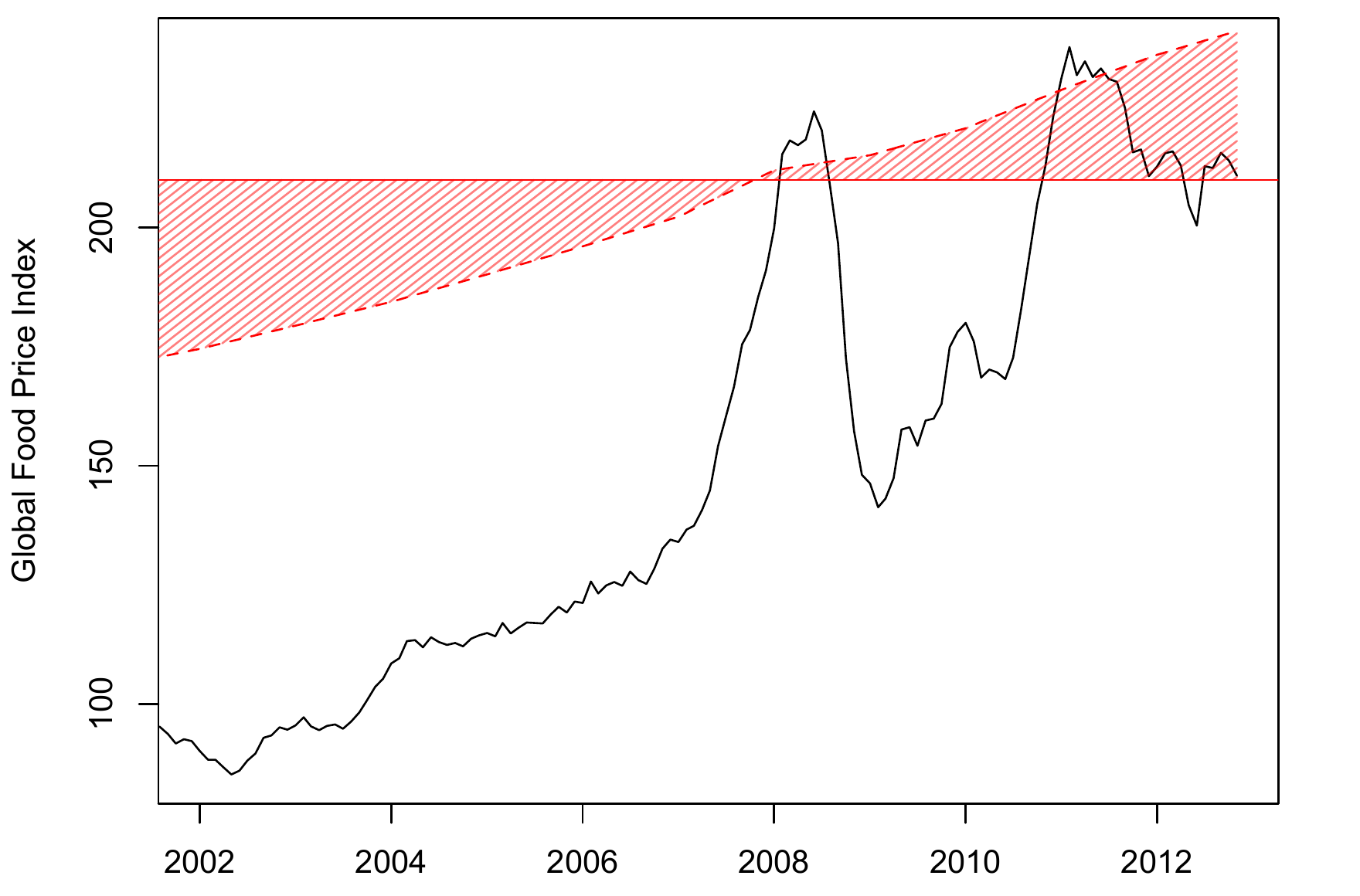}
\caption{Global food price index since 2002 \cite{indexmundi}. The threshold above which widespread food riots and revolutions occurred in 2008-08 and 2010-11 \cite{food_crises} is shown both without (solid red) and with (dashed red) inflation. 
Whether incomes of poor populations increase with inflation depends on local conditions and national policies. Since mid-2011 the food price index has hovered around the threshold value.
}
\end{figure}

This summer, a severe drought in the US interrupted a declining food price trend and contributed to a new price peak, and the effects are being felt worldwide \cite{July_update}. While the drought's impact on supply is important, our analysis reveals high levels of speculation that can amplify its impact on food prices well beyond the natural consequences of the supply shock.  Maize and wheat prices rose rapidly through July. At the beginning of August, requests for a repeal of the US ethanol mandate, which would have led to a significant increase in the food supply had it been granted, may have limited the price increases on the futures market \cite{July_update,foodbriefing,PeterTimmer}. Prices remained relatively constant from August through November. However, they remained at the threshold above which violence was found in 2008-09 and 2010-11, as seen in Figure \ref{fig1}.

\begin{figure}[tb]
\centering
\refstepcounter{figref}\label{fig2}
	\includegraphics[width=170mm]{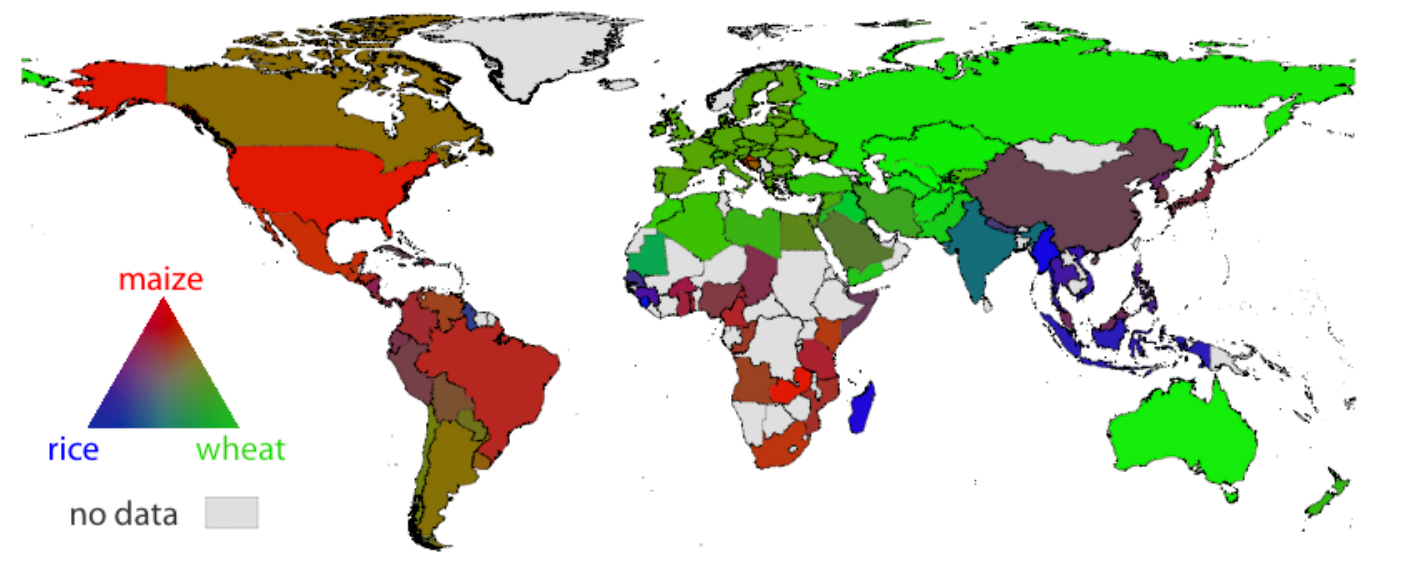}
\caption{Map showing different levels of consumption for maize (red), wheat (green) and rice (blue), by country. Map is from Ref.\ \cite{grainmapref}, based on data from Ref.\ \cite{grainmapdata}.}
\end{figure}

When prices are significantly higher than the threshold, as they were in 2007-08 and 2010-11, widespread violence can be expected. When the prices are proximate to the threshold, incidents of violence should be more sensitive to the specifics of local conditions.
Due to cultural differences, different countries rely primarily upon different grains, mostly maize, wheat or rice (Fig.\ \ref{fig2}).
Also, local prices are not always fully coupled to global prices. Finally, national policies play an important role due to food subsidies and whether salaries increase with inflation, which may be pegged more or less directly to the food prices. These considerations suggest that local conditions and the distinct impacts of different grain prices can play a pivotal role in determining the societal response to price increases. 

\begin{figure}[tb]
\centering
\refstepcounter{figref}\label{fig3}
\includegraphics[width=135mm]{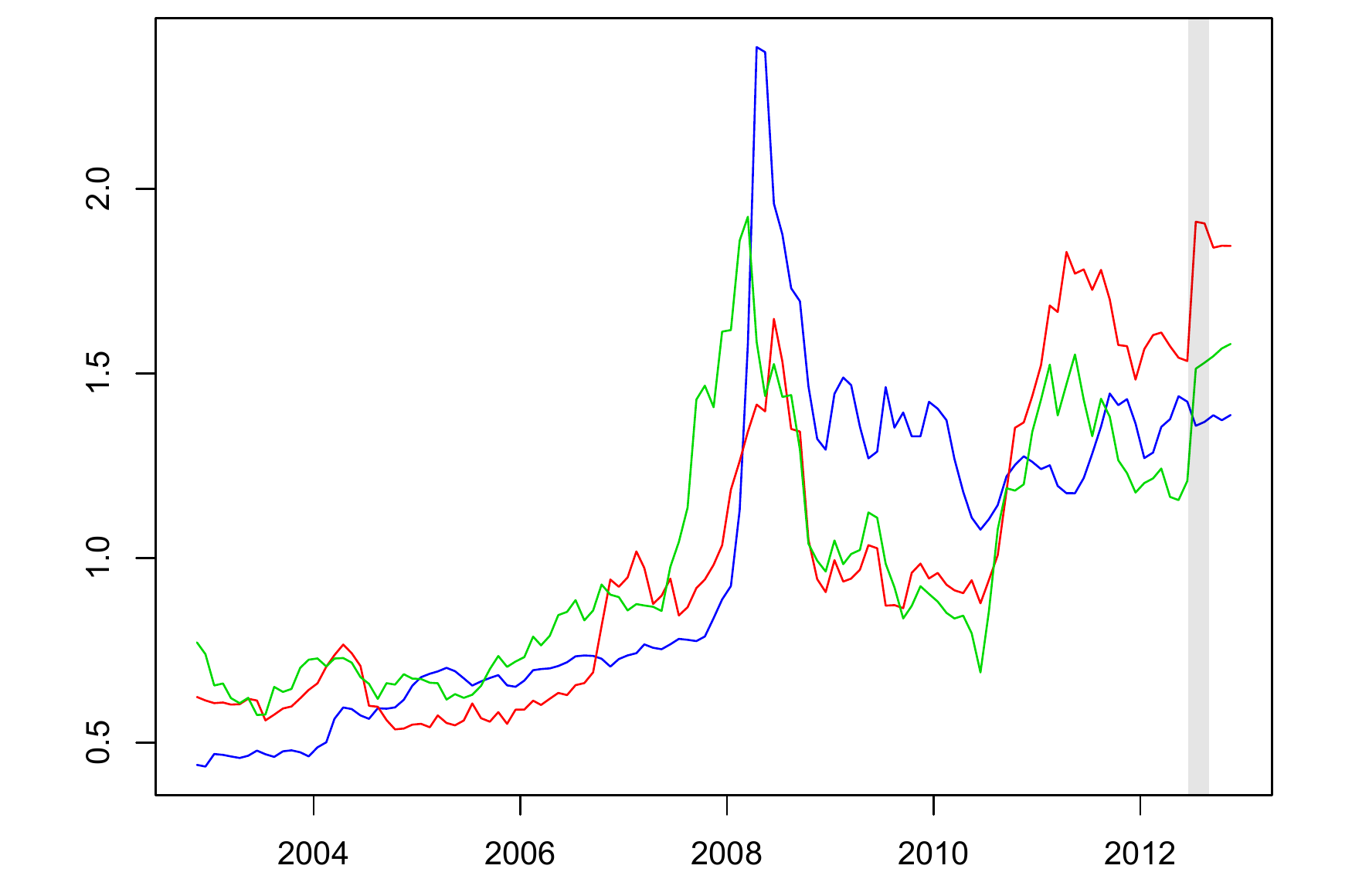}
\caption{Futures prices for maize (red), wheat (green) and rice (blue) since 2002, normalized by the average price of each grain for that time period. Shaded bar is the period of severe drought in the US in the summer of 2012.}
\end{figure}

The prices of different grains (Fig. \ref{fig3}) are partially but not completely linked, 
with rice having a more distinct behavior than maize and wheat \cite{Timmer2008}. In particular, maize prices in the summer of 2012 reached and persisted at historic highs, and wheat prices were also high. The price of rice has been relatively stable since 2009 and well below the peak reached in 2008.

\begin{figure}[tb]
\centering
\refstepcounter{figref}\label{fig4}
	\includegraphics[width=150mm]{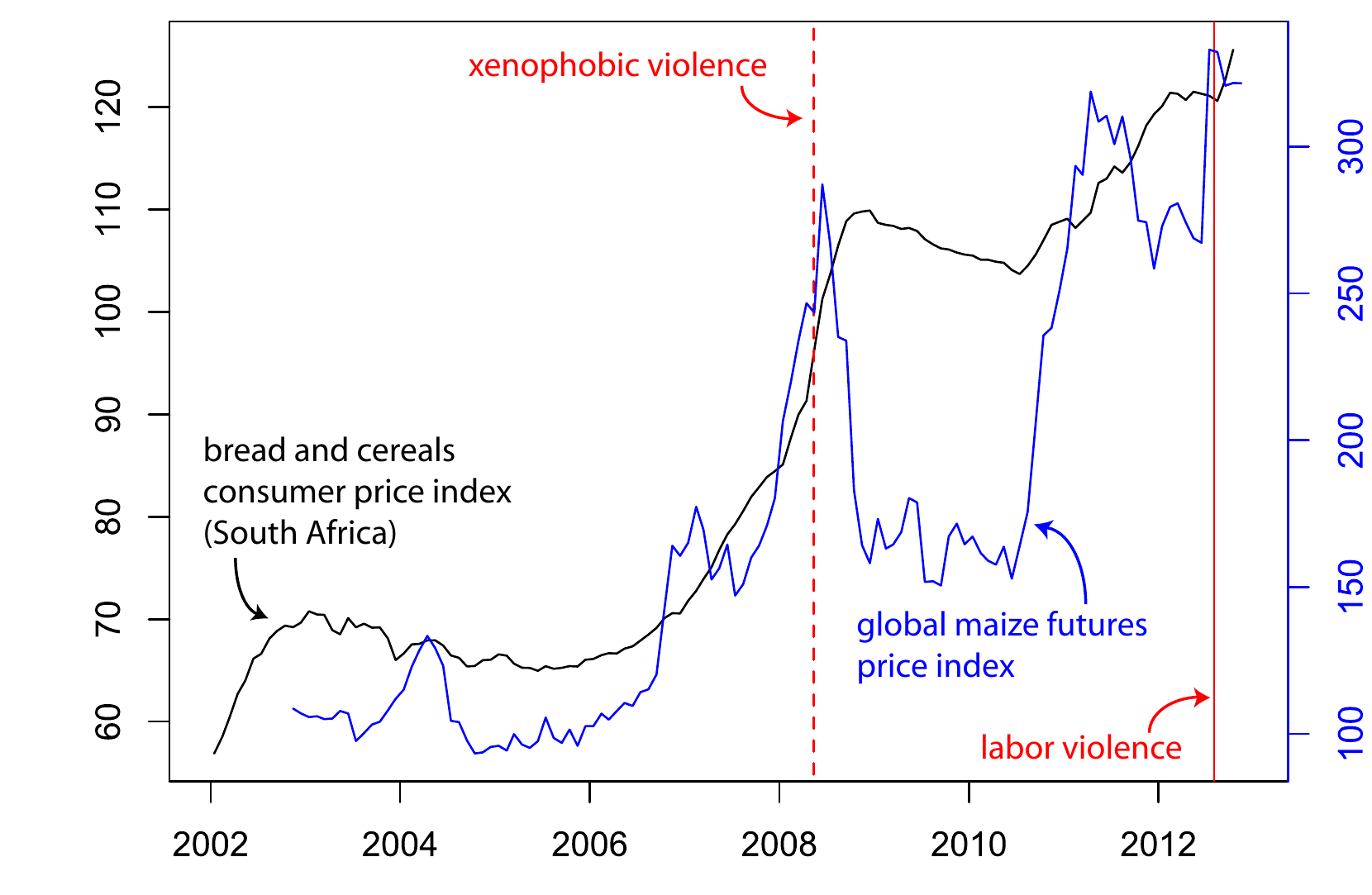}
\caption{Consumer price index for bread and cereals in South Africa since 2002 (black, left axis) \cite{statssa} and global maize prices (blue, right axis). Red solid vertical line indicates beginning of deadly riots in platinum mines, and red dashed line indicates period of severe xenophobic riots. Local prices have increased with global prices but have not correspondingly decreased when prices declined.}
\end{figure}

In South Africa, maize is the staple grain. In 2009, the most recent year of data available from the UN Food and Agricultural Organization  \cite{FAOfood},
more than half of the per capita calorie intake was from grains, of which 57\% was from maize and 32\% from wheat.
Poorer consumers are more dependent on maize and wheat and their ability to afford them is more sensitive to their prices than higher income consumers. 
The unusually violent and deadly worker riots at platinum mines starting in August of 2012 \cite{earliest,most_violent} 
coincided both with record global maize prices and record high prices for basic food items in South Africa. Figure \ref{fig4} shows the Consumer Price Index for bread and cereals for South Africa since 2002 \cite{statssa} and the date of the first deadly riot in the 
mines.

The riots have been attributed to labor issues and corruption in both the government and union organizations \cite{most_violent}. 
However, worker demands include significant wage increases amid claims that they are being paid ``hunger wages'' that do not cover basic necessities for their families \cite{hungerwages,hungerwages2}, and strikes have stopped where wage increases were granted \cite{hungerwages3}. Moreover, while labor strikes are not uncommon in South Africa, these events have been widely considered to be the most violent in the country since the end of apartheid in 1994 \cite{most_violent,massacre}. This is consistent with the view that when people are unable to feed themselves and their families, desperation leads to social unrest \cite{food_crises}. 
Previously, xenophobic riots in May of 2008 stood out as the bloodiest violence since apartheid \cite{xenophobia}. These riots coincided with food riots around the world during a previous peak of global food prices. The riots have been attributed to anger about foreigners competing for limited resources---an anger which would be exacerbated by high food prices. 

Food prices in South Africa have doubled since 2006 and the increases are strongly associated with observed periods of extreme violence. Our current results, as well as recent news of food-related protests in Haiti \cite{Haiti,Haiti2,Haiti3} and Argentina \cite{Argentina,Argentina2}, combined with our previous analysis of the role of food prices in food riots and the Arab Spring point to the importance of food prices in social unrest worldwide and the suffering of poor populations. The relatively constant prices during the fall of 2012, and the comparatively low prices of rice should not keep policymakers from recognizing that food prices are historically high. The deregulation of commodity futures markets and the diversion of almost 50\% of the US maize crop to ethanol are ill-advised policies that should be changed. Awareness of the impact of global food prices should also influence countries around the world to mitigate the impact on poor populations, including workers at the low end of the wage scale. 

We thank Peter Timmer, Dominic Albino, Karla Bertrand and Clare Fitzgerald for helpful comments on the manuscript.

\end{document}